\documentclass[%
reprint,
superscriptaddress,
groupedaddress,
runinaddress,
 amsmath,amssymb,
aps,
pra,
]{revtex4-2}

\usepackage{graphicx}
\usepackage{dcolumn}
\usepackage{bm}
\usepackage[utf8]{inputenc}
\usepackage[T1]{fontenc}
\usepackage[dvipsnames]{xcolor}

\begin{document}

\preprint{APS/123-QED}
\title{Near-unity efficiency in ridge waveguide-based, on-chip single-photon sources}
\author{Yujing Wang}
\author{Luca Vannucci}
\affiliation{
Department of Photonics Engineering, Technical University of Denmark, Ørsteds Plads, Building 343, DK-2800 Kongens Lyngby, Denmark
}
\author{Sven Burger}
\affiliation{JCMwave GmbH, Bolivarallee 22, D-14050 Berlin, Germany}

\altaffiliation[Also at ]{Zuse Institute Berlin, Takustraße 7, 14195 Berlin, Germany}
\author{Niels Gregersen}
 \email{ngre@fotonik.dtu.dk}
\affiliation{
 Department of Photonics Engineering, Technical University of Denmark, Ørsteds Plads, Building 343, DK-2800 Kongens Lyngby, Denmark
}




\begin{abstract}
We report a numerical design procedure for pursuing a near-unity coupling efficiency in quantum dot-cavity ridge waveguide single-photon sources by performing simulations with the finite element method. Our optimum design which is based on a 1D nanobeam cavity, achieves a high source efficiency $\epsilon_{xy}$ of 97.7$\%$ for an isotropic in-plane dipole, together with a remarkable Purcell factor of 38.6. Such a good performance is mainly attributed to the high index contrast of GaAs/SiO$_2$ and a careful cavity design achieving constructive interference and low scattering losses. Furthermore, we analyze the bottleneck of the proposed platform, which is the mode mismatch between the cavity mode and the Bloch mode in the nanobeam. Accordingly, we present the optimization recipe of an arbitrarily high-efficiency on-chip single-photon source by implementing a taper section, whose high smoothness is beneficial to gradually overcoming the mode mismatch, and therefore leading to a higher Purcell factor and source efficiency.  Finally, we see good robustness of the source properties in the taper-nanobeam system under the consideration of realistic fabrication imperfections on the hole variation.
\end{abstract}

\maketitle

\section{Introduction}

A deterministic and high-quality single-photon source (SPS) has a great potential to become an indispensable building block in numerous applications ranging from quantum computing \cite{Obrien2007,Thomas2021,Knill2001}, quantum cryptography \cite{Beveratos2002,Kupko2020,Schmidt2020} to large-scale on-chip quantum information processing \cite{Aharonovich2016,Kok2007,Obrien2009,Lodahl2015}. The semiconductor quantum dot (QD)-based scheme features a wide operational range, on-demand photon generation, good stability, and compatibility with on-chip technology \cite{Aharonovich2016,Kok2007,Senellart2017}, making it outstanding from the many candidates. However, the QD emission in bulk material has no directional preference, leading to an extremely low collection efficiency of less than 1\% in free space \cite{Chen2018}. Employing a nanostructure and engineering a cavity around the QD allows the guidance of photon propagation and boosts the coupling efficiency into a single guided mode thanks to Purcell-enhanced spontaneous emission (SE) \cite{Ding2016,Wang2020,Wang2019,Bleuse2011,Gregersen2016}. State-of-the-art micropillar \cite{Wang2020}, and photonic crystal (PC) waveguide cavities \cite{Lund-Hansen2008,Schwagmann2011,Arcari2014,Immo2015,Uppu2020}, can offer a near-unity source efficiency. However, the rotationally symmetric property of the micropillar SPS results in an equal excitation on two polarization orientations. Thus, the extraction efficiency cannot break through the 50\% bottleneck after the post-filter operation under the resonant excitation strategy \cite{Gur2021}. On the other side, PC-based sources suffer from poor out-coupling into single-mode waveguides, which is a limitation towards large-scale integration \cite{McNab2003}. A straightforward way to avoid these drawbacks is to embed the QD into a semiconductor ridge waveguide, which features low propagation loss \cite{Pu2016,Inoue1985} and selective guidance of photons with specific polarization orientation \cite{Stepanov2015}. Applying this strategy, the coupling efficiency
from an InAs QD to a bare GaAs ridge waveguide located on the top of a low index silica substrate has achieved $\sim$60\% \cite{Stepanov2015}. Recently, a work presents a rectangular distributed Bragg reflector (DBR) holes-based SPS featuring an efficiency of 86\% \cite{Hepp2020}. Nevertheless, realizing a near-unity efficiency comparable to the state-of-the-art micropillar or PC waveguide SPS in semiconductor ridge waveguide platform is still of critical importance and under pursuing. 

In this work, we numerically investigate the SE rate and source extraction efficiency of the InAs QD-GaAs cavity waveguide coupling system on a SiO$_2$ chip. Thanks to the Purcell effect introduced by the carefully-designed cavity, the directional improvement in power funneled into the waveguide fundamental guided mode leads to a near-unity source efficiency comparable to the state-of-the-art. Moreover, our platform shows a good tolerance with respect to realistic fabrication imperfections. In the end, we propose an optimization recipe to achieve arbitrarily high efficiency in the QD-nanobeam cavity waveguide platform.  
\begin{figure}[htbp]
\centering
{\includegraphics[width=8cm]{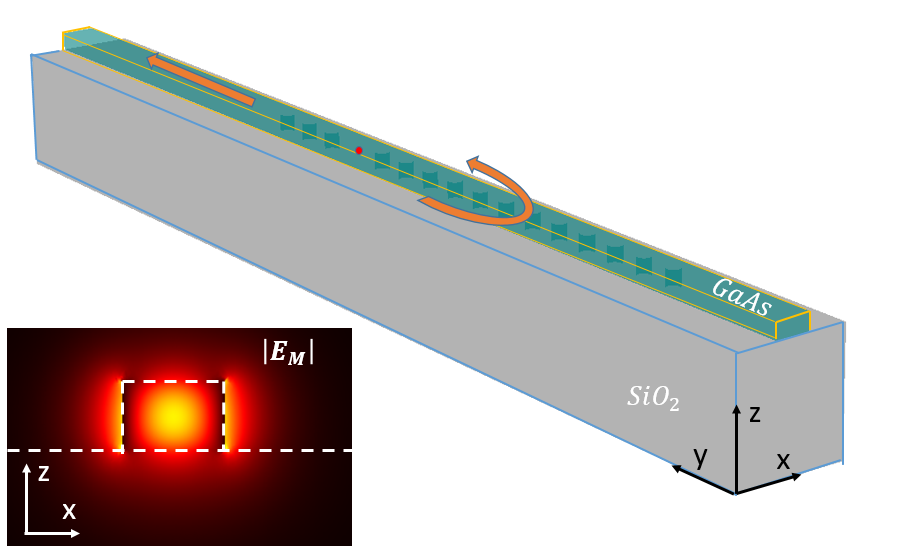}}
\caption{Sketch of the GaAs nanobeam cavity waveguide on top of the SiO$_2$ substrate. An InAs quantum dot denoted by the red circle is embedded in the center of the cavity sandwiched between two mirrors formed by cylindrical air holes. The curved arrow indicates most of the light going to the bottom mirror is reflected back to the cavity, and this part escapes from the top mirror along with the light emitting upwards, as demonstrated by the straight arrow. Inset is the 2D mode profile of the fundamental mode $M$ propagating along the bare ridge waveguide.}
\label{fig:Figure1}
\end{figure}

\section{General Concept}

The structure under consideration in this work is schematically depicted in Fig.\ \ref{fig:Figure1}. An infinitely long GaAs ridge waveguide surrounded by air is on the top of a substrate with a lower index material SiO$_2$. The Inset of Fig.\ \ref{fig:Figure1} shows that the fundamental guided mode $M$ is well confined inside the structure instead of leaking into the substrate, mainly thanks to the significant index contrast between GaAs and SiO$_2$. An InAs QD with the free-space emission wavelength of 940nm is embedded on the central axis of the waveguide. 
We build a cavity composed of two sets of DBRs formed by cylindrical air holes around the QD to improve its SE rate with the help of the Purcell effect. Considering SE wavelength, the refractive indexes are set as $n_{\rm GaAs}=3.45$, $n_{\rm SiO_2}=1.45$, and $n_{\rm Air}=1$ \cite{Stepanov2015}. We adopt a finite element method (FEM) based commercial solver (JCMsuite \cite{schneider2018numerical}) and the usage of HPC cluster hardware \cite{DTU_DCC_resource} to perform fully three-dimensional numerical simulations in this work. The semiconductor QD is modeled as an in-plane $xy$-polarized dipole under the consideration of the dipole approximation \cite{novotny2012principles,schneider2018numerical}. Correspondingly, we define $P_{x(y)}$ as the power emitted by the $x/y$-polarization configuration. The Purcell effect describes the enhancement of the QD SE rate by the nanostructure. The Purcell factor can quantify this effect and is defined as:

\begin{alignat}{1}
F_{x(y)} &= \frac{P_{x(y)}}{P_0}, 
\label{eq:Purcell}
\end{alignat}

where $P_0$ is the power emitted by the QD in the bulk material. As the figure of merit, the coupling efficiency of an $xy$-polarized dipole is given by \cite{Hoehne2019}:

\begin{alignat}{1}
\epsilon_{xy}&= \frac{P_{M,x}+P_{M,y}}{P_x+P_y}, 
\label{eq:epsilon_xy}
\end{alignat}

where $P_{M,x(y)}$ denotes the power emitted by the $x/y$-polarized dipole funneled into the mode $M$ and is evaluated by the overlap of the scattered electromagnetic field and the mode $M$ at the waveguide cross-section, which is pointed by the arrow in Fig.\ \ref{fig:Figure1}. When the dipole is on the central axis of the waveguide, it always yields $P_{M,y}=0$ \cite{Stepanov2015,Hoehne2019}. Therefore, we simplify the notation in the following by $P_M$ only referring to $P_{M,x}$. We also investigate the efficiency of the source only under the $x$-polarized configuration, defined as

\begin{alignat}{1}
\epsilon_x &= \frac{P_M}{P_x}, 
\label{eq:epsilon_x}
\end{alignat}
From the aspect of the numerical simulations in JCMsuite, we use the following solvers for solving different categories of problems in the whole work: a propagation mode solver for computing waveguide modes, a mode solver for computing Bloch modes for the cylindrical air hole unit cell, a scattering solver for computing scattered field with a dipole excitation and deducing Purcell factor, and a mode solver for computing the modes of the full structure and deducing Q factor \cite{schneider2018numerical}. 
\section{Designs and Results}
\subsection{QD coupled to a ridge waveguide}
\begin{figure*}[htbp]
\centering
{\includegraphics[width=16cm]{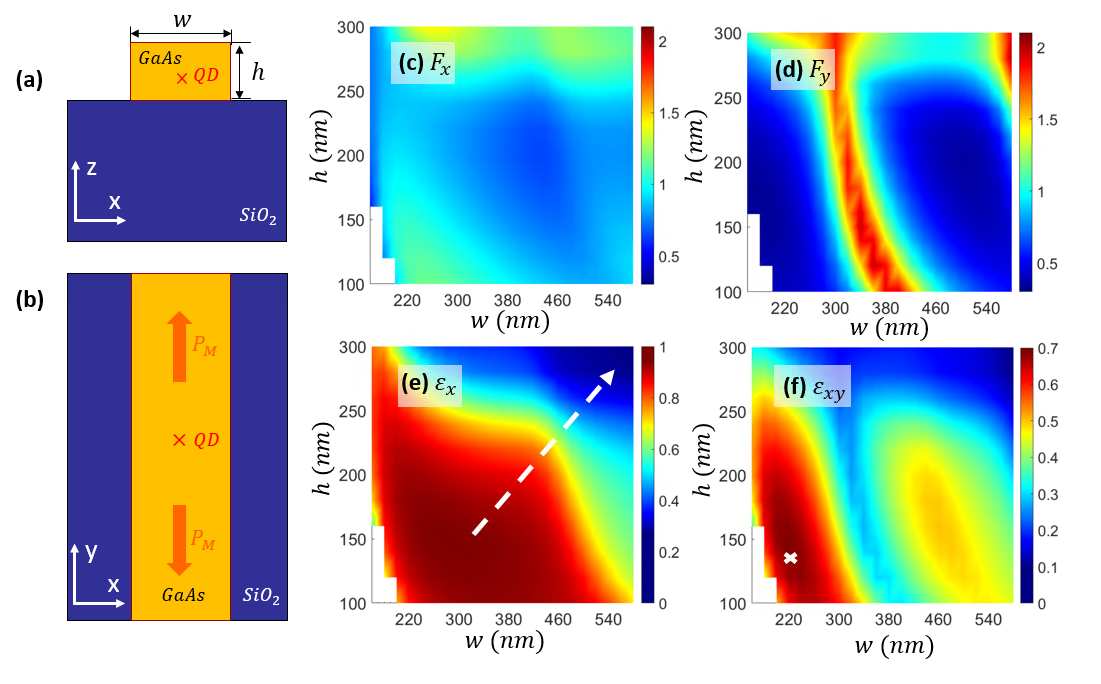}}
\caption{(a)/(b) Front view/top view of the QD-bare ridge waveguide system with physical dimensions: $w$= waveguide width, $h$= height, and material properties: substrate SiO$_2$ (dark blue), waveguide GaAs (yellow), embedded QD (red cross). $P_M$ denotes the power emitted by $x$-polarized dipole into the fundamental propagation mode $M$. In this bare waveguide, two opposite arrows indicate two equal output channels.  (c)/(d) Purcell factors $F_{x/y}$ for $x/y$-polarized dipole and (e)/(f) coupling efficiencies $\epsilon_x/\epsilon_{xy}$ for $x/xy$-polarized dipole as a function of waveguide width and height.  The white cross represents the physical dimensions yielding the optimum efficiency.} 
\label{fig:Figure2}
\end{figure*}
\begin{figure*}[htbp]
\centering
{\includegraphics[width=16cm]{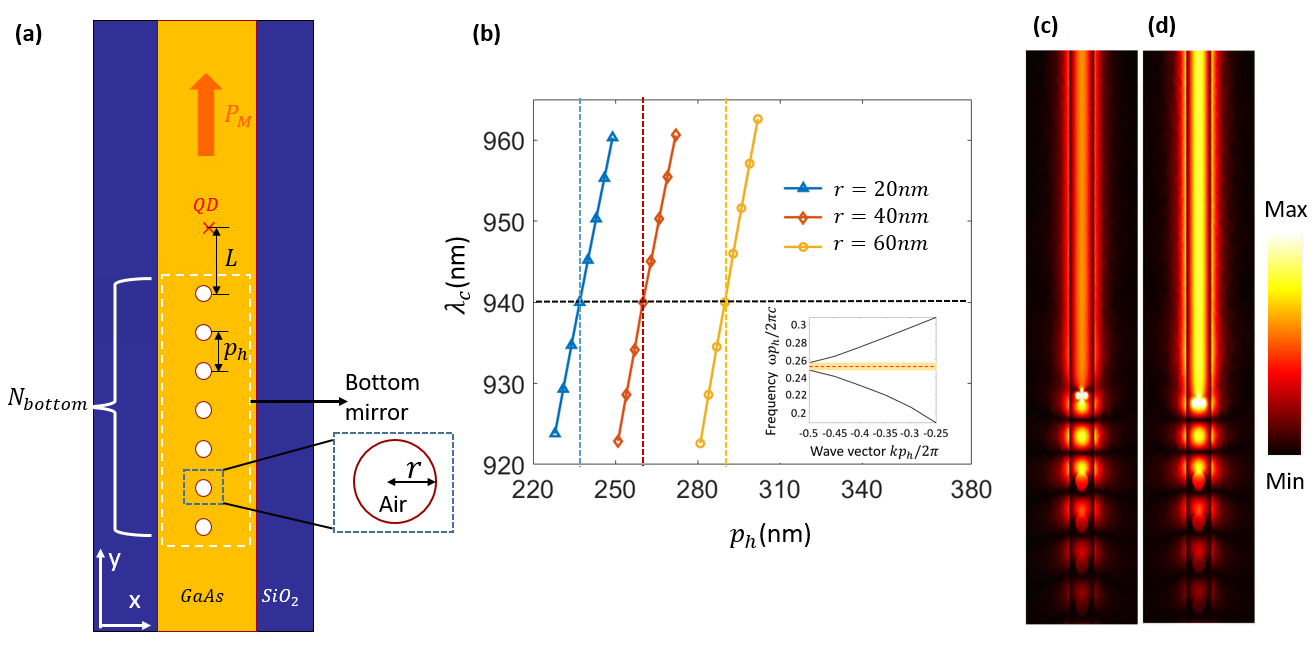}}
\caption{(a) Top view of the QD-ridge waveguide system with bottom mirror, formed by an array of periodic and uniform holes. $N_\textrm{bottom}$ denotes the number of holes in the bottom mirror, $L$ and $p_h$ is the distance from QD to the bottom mirror and between two adjacent periods. Inset shows the enlarged view of one of the air holes with the radius of $r$.  Emitted photons can only propagate through one output channel, as the arrow indicates. (b) Center wavelength $\lambda_c$, obtained from the bandgap calculation of three different hole radii ( $r=20$nm, $r=40$nm, $r=60$nm), as a function of periodicity $p_h$. The intersections of horizontal and vertical dashed lines imply three suitable $p_h$ to reach $\lambda_c$=940nm (designed wavelength). Inset shows the bandgap (yellow shade) of a period with $r=20$nm, $p_h=237$nm, where the red dashed line indicates the center frequency ($\lambda_c=940$nm). (c)/(d) Field profile $\left| {E_x}\right|$ when QD is at a random position/an antinode relative to the bottom mirror under the same color bar.}
\label{fig:Figure3}
\end{figure*}
First, we investigate the dependence of $\epsilon_{xy}$ on the waveguide geometry for an infinitely long and pristine waveguide (with no holes), whose cross-section and top view are depicted in Fig.\ \ref{fig:Figure2}(a) and (b). We consider a rigde waveguide of height $h$ and width $w$ in the ranges $h\in[100\textrm{nm},300\textrm{nm}]$, $w\in[200\textrm{nm},580\textrm{nm}]$ respectively, with a step-size $\Delta h=\Delta w=10$nm. Transparent boundary conditions are implemented using perfectly matched layers (PML) \cite{schneider2018numerical} to perform the infinitely spatial scattering simulations. Then, we compute the mode overlap and the emitted power by $x/y$- polarized dipole for each set of parameters to calculate the Purcell factors and efficiencies according to Eqs. (\ref{eq:Purcell}-\ref{eq:epsilon_x}), and the results are shown in Fig.\ \ref{fig:Figure2}(c)-(f). The white areas correspond to the specific sets where no guided mode can be found. 

Comparing Fig.\ \ref{fig:Figure2}(c) and (d), the Purcell factor of $y$-polarized dipole $F_y$ is more sensitive to the dimensional variation than that of $x$-polarized dipole $F_x$. This is because the oscillation axis of the $y$-polarized emitter ($y$-axis) is perpendicular to the waveguide cross-section ($xz$-plane) \cite{Hoehne2019}. Next, we consider the sum of the power extracted from the top and bottom outputs, $P_{M,\textrm{top}}$ and $P_{M,\textrm{bot}}$, as shown by the arrows in Fig.\ \ref{fig:Figure2}(b), when evaluating the source efficiencies. In principle, $P_{M}$=$P_{M,\textrm{top}}$+$P_{M,\textrm{bot}}$. For a purely $x$-polarized emitter, the coupling efficiency $\epsilon_x$ can reach as high as 95\%. In addition, the area with a small cross-sectional size can maintain high efficiency of over 90\%, as shown in the dark-red area in Fig.\ \ref{fig:Figure2}(e), where we can also see $\epsilon_x$ diagonally drops, as indicated by the arrow. The reason is that a wider structure supports additional guided modes, thus decreasing the relative emission into the fundamental mode. The white cross in Fig.\ \ref{fig:Figure2}(f) represents the geometric parameters $h=140$nm and $w=220$nm yielding a maximum coupling efficiency $\epsilon_{xy}$ of 70\% for an $xy$-polarized emitter. The structures that we investigate in the following will inherit this set of dimensions. 
As shown in Fig.\ \ref{fig:Figure2}(d)-(f), higher values of $\epsilon_{xy}$ coincide with high $\epsilon_x$ and low $F_y$. Moreover, the bright/dark areas in Fig.\ \ref{fig:Figure2}(f) almost perfectly correspond to the dark/bright areas in Fig.\ \ref{fig:Figure2}(d). This is due to $P_{M,y}=0$ and the strong dependence of efficiencies on the dipole orientation ($x$). Therefore, the distribution of $\epsilon_{xy}$ relies on the interplay of $\epsilon_x$ and $F_y$, and a relative suppression of $F_y$ will significantly contribute to a higher $\epsilon_{xy}$. However, a bare waveguide always features $P_{M, \textrm{top}} = P_{M, \textrm{bot}}$, which means we will always lose 50\% of the light if we only collect from the top. Therefore, in the following we will try to make $P_{M, \textrm{top}} \gg P_{M, \textrm{bot}}$ by engineering a highly reflective bottom mirror.

\subsection{Design concept and results for a DBR in a ridge waveguide}

 As depicted in Fig.\ \ref{fig:Figure3}(a), the DBR mirror is composed of cylindrical air holes with a number of $N_\textrm{bottom}$. The hole radius $r$ and the distance between successive hole centers, periodicity $p_h$, should be carefully selected to ensure a good reflection from the bottom mirror at the designed SE wavelength 940nm. We find the correct periodicity by performing a photonic band structure calculation for 3D unit cell of the PC in a 1D-periodic arrangement for different values of $p_h$ and $r$, as shown in the inset of Fig.\ \ref{fig:Figure3}(b). The yellow shade indicates the bandgap \cite{Joannopoulos2008} of the periodic structure with parameters $r=20$nm, $p_h=237$nm, in the center of which a red dashed line denotes its center wavelength $\lambda_c$ precisely at 940nm. For the highest possible reflectivity in DBR, the band gap should be engineered exactly around the SE frequency. After scanning the periodicities for three radii that we would investigate in the following and extracting all the corresponding center wavelengths, we finally get the impact of $p_h$ on $\lambda_c$, as shown in Fig.\ \ref{fig:Figure3}(b). It reveals that a longer period is necessary for a larger hole or a longer wavelength to achieve the greatest reflection.  The intersections of the dashed lines in Fig.\ \ref{fig:Figure3}(b) represent the optimal $p_h$ of three periodic structures at the designed wavelength 940nm. They are $p_h=237\textrm{nm} (r=20\textrm{nm}), p_h=260\textrm{nm} (r=40\textrm{nm}), p_h=290\textrm{nm} (r=60\textrm{nm})$, respectively. By applying these parameters, we can expect a fairly high bottom reflection to appear in this structure.

Next, a stronger Purcell enhancement can be achieved by the constructive interference between the forward-propagating photons and the ones reflected at the bottom mirror \cite{Hoehne2019}, which requires the dipole to be precisely at an antinode relative to the mirror.  We first place the emitter at a random distance $L$ to the bottom mirror to obtain a field profile, where we can observe the position of the antinodes. As shown in Fig.\ \ref{fig:Figure3}(c), there are three antinodes between the dipole and the mirror, indicated by three bright standing wave patterns. The second step is to choose one of the antinodes and relocate the emitter there. Here, we avoid the first antinode in case there is any influence induced by the close distance between QD and DBR.  In this structure, we choose the third antinode to locate the dipole, and the field profile is shown in Fig.\ \ref{fig:Figure3}(d), where the output signal is much stronger than that in Fig.\ \ref{fig:Figure3}(c). As mentioned above, this results from a perfect constructive interference when the emitter is precisely at the antinode. For different radii, the distances of QD to the bottom mirror used in this work are given as follow: $L=289\textrm{nm} (r=20\textrm{nm}), L=298\textrm{nm} (r=40\textrm{nm}), L=312\textrm{nm} (r=60\textrm{nm})$. Although we observed an enhanced SE rate and $P_{M, \textrm{top}} \gg P_{M, \textrm{bot}}$ by only implementing a bottom mirror, we are still pursuing a better Purcell factor and higher source efficiency, as shown in the next section.

\subsection{QD in nanobeam cavities with and without taper}
\begin{figure*}[htbp]
\centering
{\includegraphics[width=16cm]{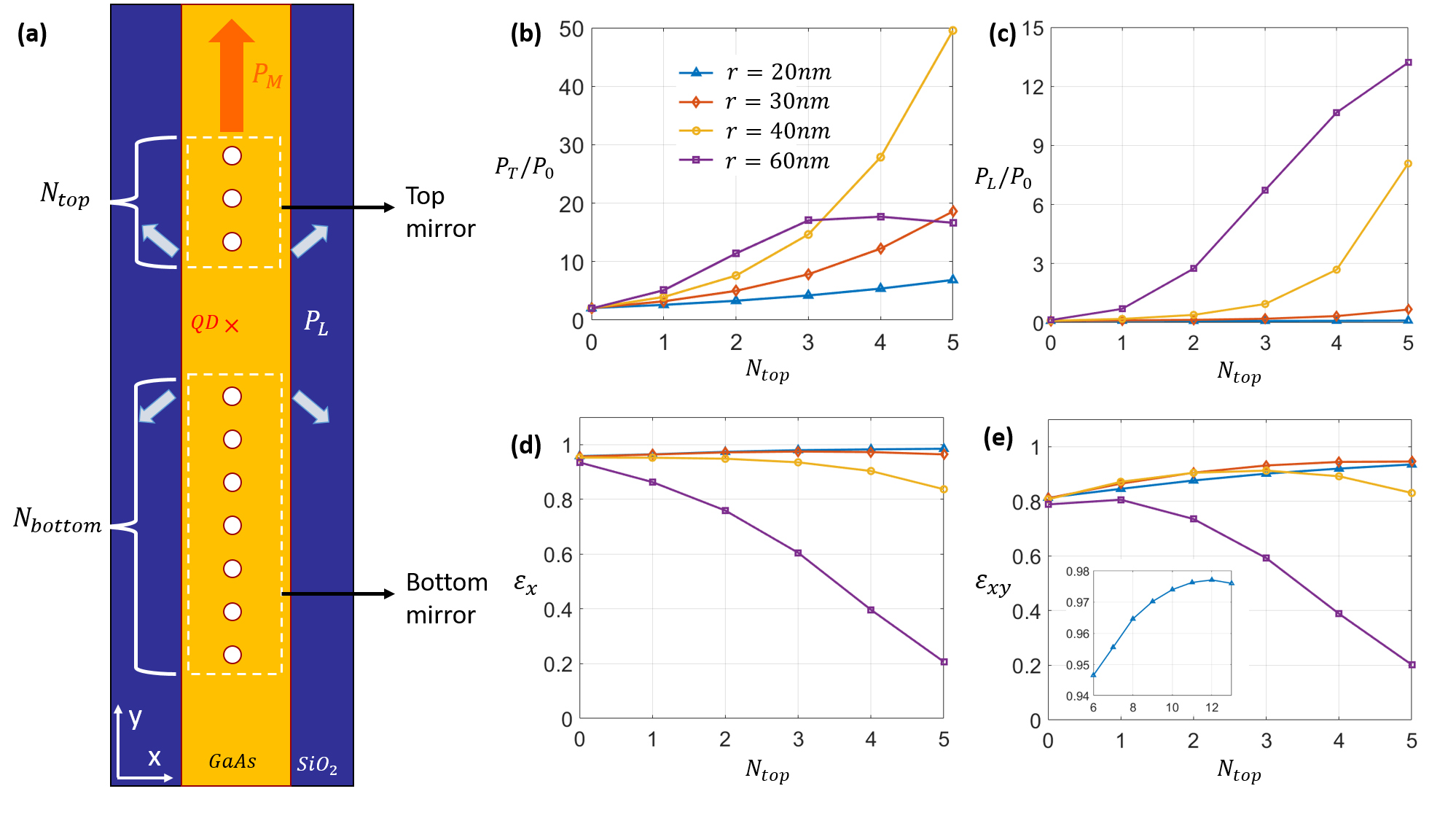}}
\caption{(a) Top view of the QD-nanobeam cavity waveguide (regular DBR) coupling system. $N_\textrm{top}/N_\textrm{bottom}$ is the number of periods in top/bottom mirror. For $x$-polarized dipole, $P_M$ denotes the power funneled into fundamental mode $M$, all the other power lost by radiation $P_L$ is depicted by the small arrows, total power $P_T=P_M+P_L$. (b)/(c) Emitted/lost power $P_{T/L}$ and (d)/(e) coupling efficiencies $\epsilon_x/\epsilon_{xy}$ for $x/xy$-polarized dipole as a function of $N_\textrm{top}$. Inset shows the maximum efficiency $\epsilon_{xy}$ can reach 97.7\% with parameters: $r=20$nm, $N_\textrm{top}=12$.}
\label{fig:Figure4}
\end{figure*}

The third step in the design procedure is to construct a complete cavity, in which two sets of DBR sandwich the QD in the center, and thus cavity length is twice as long as $L$. By applying the parameters designed in the previous step, the cavity will further enhance the QD SE through a more substantial Purcell effect. Fig.\ \ref{fig:Figure4}(a) is the sketch of a QD-nanobeam cavity waveguide coupling system. Here, we design a top mirror with a higher transmission than the bottom one, which is achieved by using less holes. For a purely $x$-polarized dipole, part of the total emitted power $P_T$ funnels into the waveguide fundamental guided mode $M$, denoted by $P_M$. The rest lost into radiation, named $P_L$, mainly appears at the interfaces of the cavity and the DBRs.

Fig.\ \ref{fig:Figure4}(b)-(e) show the impact of the number of periods in top mirror $N_\textrm{top}$ on the SE properties and coupling efficiencies under four sets of configurations. Generally, when $N_\textrm{top}$ increases, the top reflectivity increases as well. For each data point, $N_\textrm{bottom}$ is accordingly increased to ensure the bottom reflectivity stays higher. Here, we state that the parameters used in the design of $r=30$nm are $p_h=248$nm, $L=291$nm. Fig.\ \ref{fig:Figure4}(b) demonstrates the normalized $P_T$ emitted by an $x$-polarized dipole as a function of $N_\textrm{top}$. In the range of $N_{top}\in[0,5]$, the large size design ($r=60$nm) has already reached the highest Purcell factor. In contrast, the curves for 20, 30 and 40 nm are still going up but will not increase indefinitely. Due to the scattering taking place at the cavity-DBR interface, there is a trade-off between the enhancing quality factor and the scattering loss when increasing $N_\textrm{top}$, and therefore those curves will reach a maximum at a larger $N_\textrm{top}$. 
As $N_\textrm{top}$ increases, the reflectivity from the top becomes comparable to that from the bottom, and thus more light into the free space leads to a rise in $P_L$, as shown in Fig.\ \ref{fig:Figure4}(c). Here, the structure with the smallest size ($r=20$nm) yields the lowest $P_L$. Figure 4(d) and 4(e) confirm indeed that a smaller radius is beneficial for suppressing $P_L$, as both $\epsilon_x$ and $\epsilon_{xy}$ achieve much higher values in this regime.
We find a maximum efficiency $\epsilon_{xy}$ =97.7\% for $r=20$nm and $N_\textrm{top}=12$ (and a corresponding $N_\textrm{bottom}=85$), accompanied by a Purcell factor of $F_x=38.6$, and a $Q$ factor of $668$. A single computation in the parameter scan is performed with an accuracy setting yielding an estimated numerical relative error of $0.5\%$ for Q and of $0.1\%$ for $\epsilon_{xy}$. The corresponding discrete problem has a dimension of N=14836908 and is solved on with a RAM consumption of 500GB and with a computation time of roughly 4 hours, using 6 nodes on a standard workstation. The numerical error has been estimated using a numerical convergence study with gradually increasing numerical accuracy settings.  
\begin{figure*}[htbp]
\centering
{\includegraphics[width=16cm]{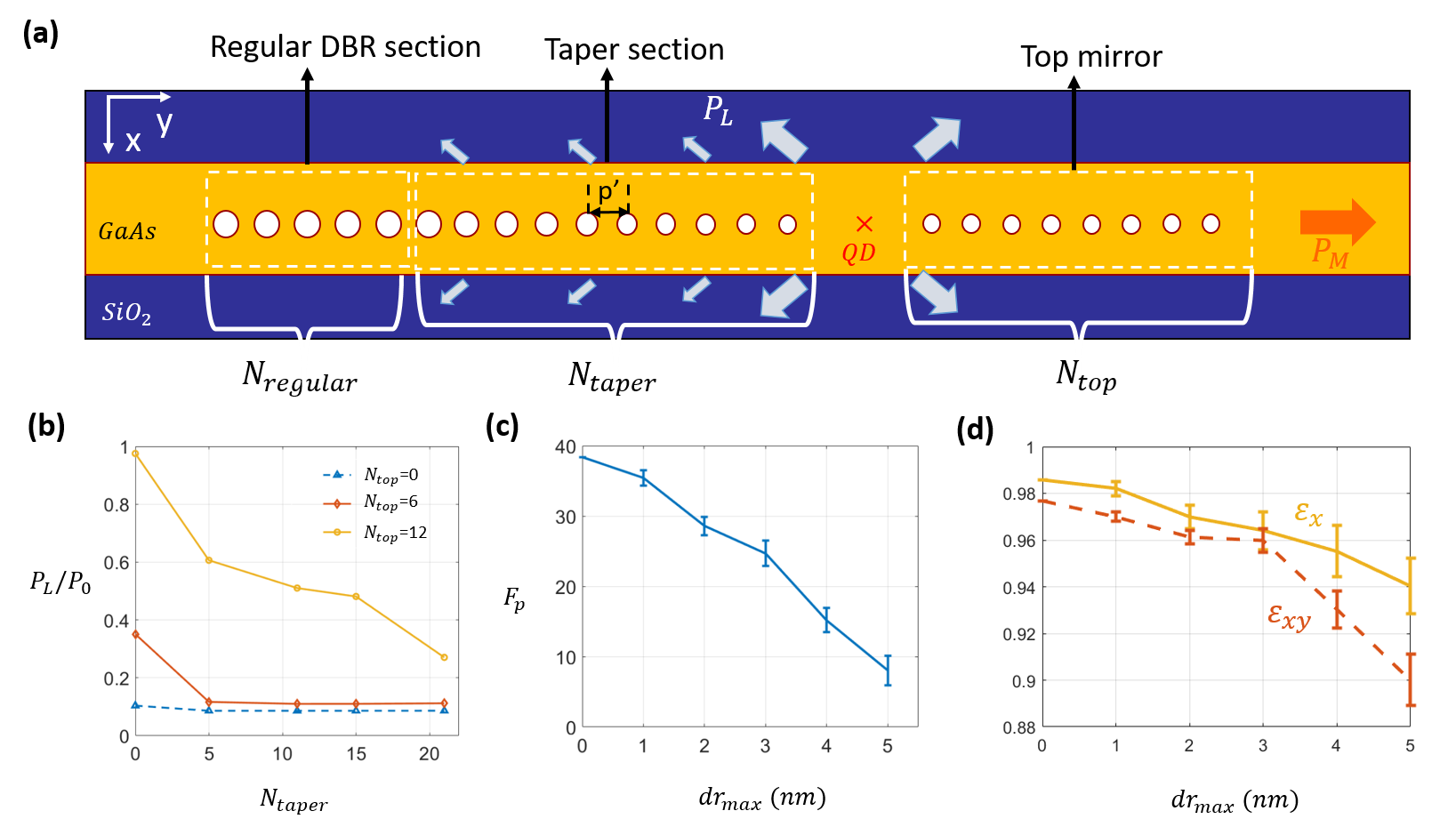}}
\caption{(a) Top view of the QD- taper cavity waveguide structure. The bottom mirror is divided into a taper section and a regular DBR section, whose number of holes are described by $N_\textrm{taper}$ and $N_\textrm{regular}$. (b) Power into radiation modes $P_L$ as a function of $N_\textrm{taper}$. (c) Purcell factor  for $x$-polarized dipole and (d) coupling efficiencies $\epsilon_x/\epsilon_{xy}$  for $x/xy$-polarized dipole as a function of maximum radius deviation $dr_{max}$ from the perfectly designed holes in the entire structure. All the data shown in (c) and (d) are obtained from 10-times simulations.}  
\label{fig:Figure5}
\end{figure*}

To make a more fabrication-friendly design, the last step in the design procedure is to find a structure with fewer periods but which can keep a near-unity coupling efficiency simultaneously. On one hand, the $r=20$nm platform effectively suppresses the scattering loss into the free space but takes a large number of periods because a single period contributes very little to the overall reflectivity. On the other hand, the $40$nm platform achieves high bottom reflection with few holes, but features stronger scattering losses due to a significant mode mismatch between the Bloch mode in the DBR mirror section and the cavity mode. One way to take advantage of both configurations is to implement a taper section \cite{Palamaru2001,Lalanne2003,Xie2021} in the bottom mirror, where the hole radius gradually increases from $20$nm in the vicinity of the QD to $40$nm. We thus consider a bottom mirror consisting of $N_\textrm{regular}$ holes of radius $40$nm and a taper section including $N_\textrm{taper}$ holes, as shown in \ref{fig:Figure5}(a), and the new variable distance between two holes $p'$ is given by $p'=(p_{h1}+p_{h2})/2$, where $p_{h1}$ and $p_{h2}$ are the individual periodicities of two adjacent periods. The principal power loss in this platform still appears at the interfaces of the cavity and two mirrors, as depicted by the thick arrows. However, the increased size in the taper section also introduces additional scattering loss, denoted by the small arrows in Fig.\ \ref{fig:Figure5}(a). We investigate the impact of  $N_\textrm{taper}$ on the normalized $P_L$, as shown in Fig.\ \ref{fig:Figure5}(b).  We can find increasing  $N_\textrm{taper}$ will effectively suppress $P_L$. In our investigations, the cavity effect is indispensable, and therefore we employ a large $N_\textrm{taper}$ in the design. Then, we propose an optimum new design yielding a coupling efficiency $\epsilon_{xy}=97.6\%$, a remarkable Purcell factor of $38.3$, and a $Q$ factor of $616$ with the parameters $N_\textrm{regular}=9$, $N_\textrm{taper}=21$,  $N_\textrm{top}=12$. Comparing with the previous result, we achieve identical performance but with only 1/3 of the total periods in the bottom mirror, contributing to a smaller mode volume and, in turn, a broader bandwidth. In principle, the figure of merit $\epsilon_{xy}$ can be further improved by using a longer taper section and smaller holes, but these come at the price of much longer structure and heavier calculations. Under the premise that the minimum design radius is 20nm, we don't implement a similar taper section in the top mirror and instead keep the regular DRB. This avoids the additional power loss introduced by the increased radius in the taper. 

\subsection{Investigation of the impact of fabrication tolerances on the performance of the SPS}

Considering the imperfections that will appear in the actual situations, the tolerance study for the optimum QD-taper cavity platform is carried out by randomly changing the radii of the well-designed holes. Fig.\ \ref{fig:Figure5}(c) and (d) demonstrate the Purcell factor and efficiencies as a function of the maximum radius deviation $dr_{max}$ between the designed radii and the possible imperfections in fabrication, which means that each hole radius is changed by an individual random amount in the interval (0, $dr_{max}$). Even under a 5nm deviation, our platform is still robust on $\epsilon_x$ and $\epsilon_{xy}$, keeping the performances of over 90\% with a standard deviation of 1\%. However, the Purcell factor is more sensitive to imperfect situations. This is because a misalignment of the new antinode with respect to QD position leads to poor constructive interference.

\section{Conclusion} 

In summary, we adopt the finite element method to numerically investigate the QD spontaneous emission and its coupling efficiency to the waveguide fundamental guided mode in a nanobeam cavity SPS. A careful cavity design can not only improve the dipole emission but also boost the source efficiency via lower scattering losses. By investigating the impact of the hole radius on the dipole emission properties, we find that the smaller the hole, the better it controls the scattering loss at the cavity-DBRs interfaces and that in the taper. Our optimum platform appears with a near-unity efficiency (97.7\%), a remarkable Purcell factor of 38.6, and a $Q$ factor of $668$. We also propose a taper to strongly reduce the number of holes in the bottom mirror by a factor of $\sim3$, and to achieve a broader bandwidth, corresponding to a slightly lower $Q$ factor of $616$. Moreover, the Purcell factor and efficiency are still comparable to those in the optimum system. Considering the feasibility in fabrication, the smallest hole applied in the current design is with a radius of 20nm. However, it is possible to pursue a source efficiency arbitrarily close to unity by reducing the radii in the overall structure and increasing the number of holes in the taper section based on our design. Finally, the platform under investigation is proven to be robust when considering imperfections in realistic fabrication processing. Therefore, our work is promising to allow for the fabrication of the nanobeam single photon sources with significantly improved performance compared to today, and provide the potential contribution to the integrated on-chip, and large-scale quantum information processing.  

\section{Acknowledgements} 
The authors acknowledge support from the European Research Council (ERC-CoG "UNITY",grant 865230), from the European Union’s Horizon 2020 Research and Innovation Programme under the Marie Sklodowska-Curie Grant Agreement No. 861097, and from the Independent Research Fund Denmark (grant DFF-9041-00046B), from the EMPIR programme co-financed by the Participating States and by the European Union’s Horizon 2020 research and innovation programme (project 20FUN05 "SEQUME" ).

\bibliographystyle{unsrt}
\bibliography{manuscript}

\end{document}